\documentclass[journal,twoside,web]{ieeecolor}
\usepackage{tmi}
\usepackage{cite}
\usepackage{amsmath,amssymb,amsfonts}
\usepackage{algorithmic}
\usepackage{graphicx}
\usepackage{textcomp}
\usepackage{dsfont}
\usepackage{color, soul}
\usepackage{diagbox}
\usepackage{makecell}

\def\BibTeX{{\rm B\kern-.05em{\sc i\kern-.025em b}\kern-.08em
    T\kern-.1667em\lower.7ex\hbox{E}\kern-.125emX}}
\usepackage{graphicx,amssymb,mathrsfs,amsmath,amsfonts,dsfont}
\usepackage{hyperref}
\usepackage{lineno}
\usepackage{stfloats}
\usepackage{algorithm}
\usepackage{algorithmic}
\usepackage{amsbsy}
\usepackage[mathscr]{euscript}
\usepackage{bm}
\usepackage{multirow}
\usepackage[utf8]{inputenc}

\newtheorem{thm}{Theorem}[section]

\newtheorem{rmk}{Remark}[section]

\begin{document}
\title{Matrix Completion-Informed Deep Unfolded Equilibrium Models for Self-Supervised $k$-Space Interpolation in MRI}
\author{Chen Luo, Huayu Wang, Taofeng Xie, Qiyu Jin, Guoqing Chen, Zhuo-Xu Cui, Dong Liang, \IEEEmembership{Member, IEEE}
\thanks{This work was supported in part by the National Key R$\&$D Program of China (2021YFF0501503, 2020YFA0712202 and 2022YFA1004202); National Natural Science Foundation of China (U21A6005, 62125111, 12026603, 62206273, 61771463, 81830056, U1805261, 81971611, 61871373, 81729003, 81901736); Key Laboratory for Magnetic Resonance and Multimodality Imaging of Guangdong Province (2020B1212060051).}
\thanks{Corresponding author: zx.cui@siat.ac.cn and dong.liang@siat.ac.cn}
\thanks{C. Luo and H. Wang contributed equally to this work}
\thanks{C. Luo, H. Wang, T. Xie, Q. Jin and G. chen are with School of Mathematical Sciences, Inner Mongolia University, Hohhot, China.}
\thanks{Z.-X. Cui and D. Liang are with Research Center for Medical AI, Shenzhen Institutes of Advanced Technology, Chinese Academy of Sciences, Shenzhen, China.}}

\maketitle

\begin{abstract}

Recently, regularization model-driven deep learning (DL) has gained significant attention due to its ability to leverage the potent representational capabilities of DL while retaining the theoretical guarantees of regularization models. However, most of these methods are tailored for supervised learning scenarios that necessitate fully sampled labels, which can pose challenges in practical MRI applications. To tackle this challenge, we propose a self-supervised DL approach for accelerated MRI that is theoretically guaranteed and does not rely on fully sampled labels. Specifically, we achieve neural network structure regularization by exploiting the inherent structural low-rankness of the $k$-space data. Simultaneously, we constrain the network structure to resemble a nonexpansive mapping, ensuring the network's convergence to a fixed point. Thanks to this well-defined network structure, this fixed point can completely reconstruct the missing $k$-space data based on matrix completion theory, even in situations where full-sampled labels are unavailable. Experiments validate the effectiveness of our proposed method and demonstrate its superiority over existing self-supervised approaches and traditional regularization methods, achieving performance comparable to that of supervised learning methods in certain scenarios.

\end{abstract}


\section{Introduction}
\label{sec:introduction}
\IEEEPARstart{M}{agnetic} resonance imaging (MRI) has emerged as a prominent clinical diagnostic tool due to its advanced capabilities in producing high-resolution images that offer exceptional visualization of soft-tissue contrast. Moreover, MRI is non-invasive and devoid of radiation, minimizing its impact on patients' well-being. However, the relatively slow data acquisition speed of MRI can lead to patient discomfort and motion artifacts. This concern has spurred increased attention towards accelerated MRI aimed at reconstructing accurate MR images from undersampled $k$-space data \cite{liang1992constrained}.

In recent years, deep learning (DL) has received extensive attention in accelerated MRI. In particular, regularization model-driven DL, which unfolds the iterative algorithms used for solving regularization models into deep neural networks and lets the model drive network structure design, has emerged as a significant focus \cite{Yang2020ADMMNet, Han2020kspace, Liang2020, Huang2021Deep}. This approach not only harnesses the potent representational capabilities of DL but also retains the interpretability of the model \cite{8962949}. From a theoretical perspective, drawing inspiration from deep equilibrium models (DEQs) \cite{2019Deep,2020Multiscale}, recent works \cite{Gilton2021Deep,10177777} have constrained the network structure to resemble a non-expansive mapping, ensuring the network's convergence to a fixed point. Furthermore, \cite{cui2022deep,wang2023convex} utilize input convex networks \cite{pmlr-v70-amos17b} to measure the distance from the network's output to the real data manifold. This compels the network's output solution to lie within the intersection of the real data manifold and the MRI forward model solution space, ensuring complete reconstruction under certain conditions.

However, these methodologies often rely on fully sampled data as labels during supervised learning. In practice, obtaining fully sampled labels within imaging scenarios tends to be challenging or unfeasible. Consequently, self-supervised neural networks have been developed to train on undersampled data itself to reconstruct fully sampled images, addressing this issue. However, there remains uncertainty about whether solely relying on undersampled data can effectively train a network to reconstruct fully sampled images in the absence of fully sampled labels. Although some attempts have been made to bridge this gap, for example, \cite{jin2019selfsupervised, yaman2021zeroshot, yaman2020self, hu2021self, kim2019loraki}, which has extended regularization model-driven deep learning methods to a self-supervised manner to enhance imaging interpretability. Unfortunately, to the best of our knowledge, there is currently no work in the self-supervised learning context that guarantees complete reconstruction of missing $k$-space data.

Rethinking traditional methods for accelerated MRI, we draw upon the principles of compressed sensing (CS) \cite{Donoho2006Compressed,Candes2006Robust} and matrix completion (MC) \cite{5454406} theory, which, under certain conditions, enable the complete reconstruction of MRI images or $k$-space data even when fully sampled MRI images or $k$-space data are unavailable. This motivates us to design a novel self-supervised learning mechanism that, under specific conditions, adheres to the principles of CS or MC theory, thus ensuring complete reconstruction in the absence of fully sampled labels for supervision.

\subsection{Contributions}
Motivated by the aforementioned issues, this paper proposes an MC-informed DL model for Self-Supervised $k$-space Interpolation in MRI with theoretical guarantees. Specifically, the main contributions of this work are summarized as follows:
\begin{enumerate}
\item Inspired by the structural low-rankness of $k$-space data and guided by the principle of Hankel matrix multiplication being equivalent to convolution, we have devised a convolutional neural network structure that adheres to the low-rankness. In other words, we have implemented regularization within the network structure itself. Furthermore, we have imposed additional constraints on the network architecture to ensure nonexpansiveness, thereby guaranteeing the network's convergence to a fixed point.
    
\item  Thanks to the well-defined network structure, based on the principles of matrix completion theory, even in self-supervised learning scenarios where full-sampled labels are lacking, under certain conditions, the fixed point to which the network converges is the unique true solution to the imaging inverse problem. In other words, the network can achieve complete reconstruction. To the best of our knowledge, this is the first self-supervised learning method proposed to achieve full reconstruction.

\item  Experiments validate the effectiveness of our proposed method and demonstrate its superiority over existing self-supervised approaches and traditional regularization methods, achieving performance comparable to that of supervised learning methods in certain scenarios.
\end{enumerate}

The paper is structured as follows. Section \ref{sec: related work} provides a review of related work, while Section \ref{sec: notations} presents notations and preliminaries. In Section \ref{sec: method}, we discuss our proposed $k$-space interpolation method and its corresponding theoretical guarantees. Implementation details are provided in Section \ref{sec: implementation}, followed by experimental results on several datasets in Section \ref{sec: experimentation}. We present our discussions in Section \ref{sec: discussion} before concluding with some remarks in the final section \ref{sec: conclusion}.

\section{Related Work}\label{sec: related work}
\subsection{Regularization Model-Driven Deep Learning}
In recent years, regularization model-driven DL, which unfolds the iterative algorithms used for solving regularization models into deep neural networks  has gained significant attention. Existing methods \cite{Han2020kspace, Liang2020, Huang2021Deep} predominantly focus on designing a regularization term learned by the network. For example, ISTA-Net \cite{Zhang2018ISTANet} learns a sparse transformation in the regularizer and integrates it into the ISTA algorithm. MoDL \cite{Aggarwal2019MoDL} learns an estimator of noise and alias patterns as regularization. ADMM-CSNet \cite{Yang2020ADMMNet} directly learns the regularization itself, although it depends on the ADMM optimization algorithm to iteratively approximate the optimal value. Another category of methods considers the generalization of $k$-space SLR regularization models into neural networks. DeepSLR \cite{Pramanik2020DeepSLR} leverages the equivalence between Hankel matrix multiplication and convolution, effectively extending the structural low-rank (SLR) regularization models into deep convolutional neural networks. Compared to image-domain model-driven DL, $k$-space model-driven DL offers more interpretability when it comes to convolution network generalization.

From a theoretical perspective, recent works \cite{2019Deep,2020Multiscale} have constrained the network structure to resemble a non-expansive mapping, ensuring the network's convergence to a fixed point. Furthermore, \cite{cui2022deep,wang2023convex} train  input convex networks to compel the network's output solution to lie within the intersection of the real data manifold and the MRI forward model solution space, guaranteeing complete reconstruction under certain conditions.
However, these methodologies often depend on fully sampled data as labels during supervised learning. Once labels are unavailable, the aforementioned theoretical results for complete reconstruction no longer hold. While \cite{jin2019selfsupervised, yaman2021zeroshot, yaman2020self, hu2021self, kim2019loraki} attempted to extend the model-driven approach to a self-supervised learning manner, to the best of our knowledge, there is currently no theoretical guarantee that networks trained solely on undersampled data can achieve complete reconstruction of missing $k$-space data.

\subsection{Structural Low-Rank $k$-Space Interpolation}
Interpolating missing $k$-space data has been intensively studied for a long time, dating back to the 1980s \cite{4307762}, and has seen significant development \cite{Griswold2002Generalized,lustig2010spirit}. Among these methods, one category primarily relies on the duality relationship between the SLR property of $k$-space data and the prior knowledge of multi-coil MR images. To be more specific, LORAKS \cite{lee2016acceleration} translates the smoothness of phase into $k$-space SLR, while ALOHA \cite{lee2016acceleration} converts the sparsity of MR image under gradient transform and smoothness of coil sensitivity into $k$-space Hankel SLR through domain dual transformation.

It is worth noting that the problem of recovering missing $k$-space data using SLR regularization can theoretically be reduced to an MC problem. Consequently, under specific conditions, SLR regularization can ensure the complete reconstruction of missing $k$-space data. This is also the impetus behind our investigation into self-supervised DL for MR reconstruction in the $k$-space domain. In this paper, we will develop a self-supervised DL within the SLR regularization paradigm, guided by the principle that Hankel matrix multiplication is equivalent to convolution. This enables us to achieve complete reconstruction of missing $k$-space data, even in situations where full-sampled labels are unavailable.
\section{Notations}
\label{sec: notations}
In this paper, matrices and vectors are all represented by bold lower case letters, i.e. $\mathbf{x}$, $\mathbf{y}$. In addition, $\mathbf{x}_{i}$ and $\mathbf{x}_{i,j}$ respectively correspond to the $i$-th column and $(i,j)$-th entry of the matrix $\mathbf{x}$, and $x_i$ denotes the $i$-th element of vector $\mathbf{x}$. 
The notation $\mathcal{F}$ or superscript $\hat{\cdot}$ denotes the Fourier transform, as shown as
\begin{equation}
\hat{\mathbf{x}}(k) = \mathcal{F}(\mathbf{x}) = \int \mathbf{x} (r) e^{-j2\pi k\cdot r} d r.
\end{equation}

This paper encompasses a range of matrix norms. The Euclidean inner product between two matrices is defined as $\langle \mathbf{x},\mathbf{y} \rangle = \text{Tr}(\mathbf{x}^H\mathbf{y})$, with the corresponding Euclidean norm commonly referred to as the Frobenius norm $\|\mathbf{x}\|_F$. The Frobenius norm is derived as $\|\mathbf{x}\|_F:=\langle\mathbf{x},\mathbf{x}\rangle$.
Without specific notation, for matrices or operators, $\|\cdot\|$ represents the spectral norm, and for vectors, $\|\cdot\|$ represents the $\ell_2$-norm.

We hereby review the relational expression connecting matrix convolution operations with Hankel matrices. For the sake of succinctness, we confine our discussion to the 1-D scenario, with extensions to higher dimensions deferred to \cite{ye2018deep}. Let vector $\mathbf{x}=[x_1,\ldots,x_n]^{T}\in\mathbb{C}^n$ and $\mathbf{s}=[s_1,\ldots,s_d]^{T}\in\mathbb{C}^d$. Define $\overline{\mathbf{s}}$ as the time-reversed version of $\mathbf{s}$, i.e., $\overline{\mathbf{s}}[i]=\mathbf{s}[-i]$.
The output of the convolution between matrix $\mathbf{x}$  and filter $\overline{\mathbf{s}}$ can be represented as a matrix multiplication:
\begin{equation}
\mathbf{x} \circledast \mathbf{s} = \mathcal{H}(\mathbf{x},d)\overline{\mathbf{s}}
\end{equation}
Here, $\circledast$ represents the convolution operator, $\mathcal{H}(\mathbf{x},d)$ corresponds to the wrap-around Hankel matrix. 
The frequently employed mathematical symbols in this paper, along with their detailed explanations, are presented in Table \ref{table: notations}.

\begin{table}[!t]
\caption{Summary of mathematical notions and corresponding notations.}
\centering
  \footnotesize
      \begin{tabular}{c|c}
        \hline \hline Notation & Notion \\
        \hline  
        $\mathbf{x}$ & multi-channel MR image \\
        $\hat{\mathbf{x}}$ & $k$-space data, $\mathbf{\hat{x}}=\text{FFT}(\mathbf{x})$ \\
        $\text{csm}_i$ & $i$-th coil sensitivity of image $\mathbf{x}$ \\
        $\Omega$ & a set of undersampled location indices \\
        $M_{\Omega}$ & mask indicating the undersampling operator \\
        $\mathbf{y}_{\Omega}$ & undersampled $k$-space data, $\mathbf{y}_{\Omega} = M_{\Omega} \hat{\mathbf{x}}$ \\
        $\mathbf{x}^H$ & Hermitian transpose of $\mathbf{x}$, i.e., conjugate and transpose \\
        $\mathcal{H}(\mathbf{x},d)$ & wrap-around Hankel matrix of $\mathbf{x}$ \\
        $\circledast$ & convolution operator \\
        \hline \hline
    \end{tabular}
\label{table: notations}
\end{table}

\section{Methodology and Theory }\label{sec: method}
In this section, we will start by introducing the process of driving a deep neural network from a $k$-space SLR model to achieve complete reconstruction with theoretical guarantees. Subsequently, we will explore the techniques for self-supervised training of this network in scenarios where full-sampled labels are not accessible. Finally, we will discuss the methods for generalizing this network to enhance its representational capabilities.

\subsection{ Structural Low-rank Model Unfolded Networks}
The multichannel $k$-space data acquisition in MRI can be mathematically represented as a forward model: 
\begin{equation}\label{eq:2}
\mathbf{y}_{\Omega} = M_{\Omega}\hat{\mathbf{x}} + \mathbf{n},
\end{equation}
where $\hat{\mathbf{x}} = [\hat{\mathbf{x}}_{1}, \ldots, \hat{\mathbf{x}}_{N_{c}}]$ and $ \mathbf{y}_{\Omega} \in \mathbb{C}^{N1\times N2\times N_{c}}$ denote the multi-channel fully sampled and undersampled $k$-space data. $N_{c} \ge 1$ is the number of coils and $\mathbf{n}$ indicates the noise in the measurement. The $j$-th element of $M_{\Omega}$ has a value of $1$ if its position index is in the undersampled set $\Omega$ and $0$ otherwise.

Reconstructing missing data from the undersampled data, i.e., solving the inverse of model (\ref{eq:2}), is an inverse problem. Therefore, it is necessary to construct a regularization model based on prior information. As mentioned in the previous section, multi-channel $k$-space data satisfies the structural low-rank prior. Hence, the interpolation of missing $k$-space data can be formulated as:
\begin{equation}\label{eq:slr}
\min_{\hat{\mathbf{x}}} \|\mathcal{H}(\hat{\mathbf{x}},d)\mathbf{s}\|_F^2~\text{s.t.}~M_{\Omega}\hat{\mathbf{x}} = \mathbf{y}_{\Omega}
\end{equation}
where $\mathbf{s}:=[\mathbf{s}_1,\ldots,\mathbf{s}_r]$ is the null space filter of the matrix $\mathcal{H}(\hat{\mathbf{x}},d)$, which can usually be estimated from the $k$-space calibration region. If the optimization objective above equals 0, it implies that the rank of structural matrix $\mathcal{H}(\hat{\mathbf{x}},d)$ is less than or equal to $r$. Therefore, equation (\ref{eq:slr}) represents the search for a structural low-rank solution that satisfies $M_{\Omega}\hat{\mathbf{x}} = \mathbf{y}_{\Omega}$.
Leveraged the equivalence between Hankel matrix multiplication and convolution, model (\ref{eq:slr}) can be transformed into 
\begin{equation}\label{eq:slr2}
\min_{\hat{\mathbf{x}}} \|\text{Conv}_{\overline{\mathbf{s}}}(\hat{\mathbf{x}})\|_F^2~\text{s.t.}~M_{\Omega}\hat{\mathbf{x}} = \mathbf{y}_{\Omega}
\end{equation}
where $\text{Conv}_{\overline{\mathbf{s}}}(\hat{\mathbf{x}}):=[ \hat{\mathbf{x}}\circledast \overline{\mathbf{s}}_1,\ldots,\hat{\mathbf{x}}\circledast \overline{\mathbf{s}}_r]$. In general, we can use the projected gradient descent (PGD) to solve constraint optimization problem (\ref{eq:slr2}) efficiently, i.e.,
\begin{equation}\label{eq:7}
    \left\{\begin{aligned}
    \mathbf{r}^{k+1} &= \hat{\mathbf{x}}^{k} - \eta \text{Conv}_{\overline{\mathbf{s}}} ^{H}\text{Conv}_{\overline{\mathbf{s}}}(\hat{\mathbf{x}}^{k})\\
    \hat{\mathbf{x}}^{k+1} &= \mathcal{P}_{\{ \hat{\mathbf{x}}|M_{\Omega}\hat{\mathbf{x}} = \mathbf{y}_{\Omega}\}} (\mathbf{r}^{k+1})
    \end{aligned}\right. 
\end{equation}
where $\eta$ is the stepsize and $\mathcal{P}_{\{ \hat{\mathbf{x}}|M_{\Omega}\hat{\mathbf{x}} = \mathbf{y}_{\Omega}\}}$ represents the projection on $\{\hat{\mathbf{x}}| M_{\Omega}\hat{\mathbf{x}} = \mathbf{y}_{\Omega}\}$. Clearly, (\ref{eq:7}) represents a recursive residual convolutional neural network with kernel $\overline{\mathbf{s}}$. Thus, we have successfully implemented the design of the convolutional network structure driven by the $k$-space SLR regularization model.

For simplicity, we abbreviate $\mathcal{G}_{\mathbf{s}}(\cdot):=\cdot-\eta \text{Conv}_{\overline{\mathbf{s}}} ^{H}\text{Conv}_{\overline{\mathbf{s}}}(\cdot)$ and  $\mathcal{P}:=\mathcal{P}_{\{ \hat{\mathbf{x}}|M_{\Omega}\hat{\mathbf{x}} = \mathbf{y}_{\Omega}\}}$. Next, we will study whether executing (\ref{eq:7}) can converge to the optimal solution of (\ref{eq:slr2}) through the following theorem.
\begin{thm}\label{thm:1}
 Suppose  $\lambda_{\max}( \text{Conv}_{\overline{\mathbf{s}}} ^{H}\text{Conv}_{\overline{\mathbf{s}}}) \leq 1-\epsilon$ with $\epsilon<1$. When $\eta<1/(1-\epsilon)$, as the iterations progress, the sequence generated by algorithm (\ref{eq:7}) will converge to a fixed point, that is, $\hat{\mathbf{x}}^{\infty}= \mathcal{P}(\mathcal{G}_{\mathbf{s}}(\hat{\mathbf{x}}^{\infty}))$. Furthermore, this fixed point is the optimal solution for (\ref{eq:slr2}) or (\ref{eq:slr}).
\end{thm}
The proof is shown in the appendix. 
\begin{rmk}
The above theorem requires that the maximum eigenvalue of $\text{Conv}_{\overline{\mathbf{s}}} ^{H}\text{Conv}_{\overline{\mathbf{s}}}$, denoted as $\lambda_{\max}(\text{Conv}_{\overline{\mathbf{s}}} ^{H}\text{Conv}_{\overline{\mathbf{s}}})$, is less than $1-\epsilon$. In practice, we can achieve this by spectral normalization \cite{miyato2018spectral}.  
\end{rmk}

Since the full-sampled $k$-space data $\hat{\mathbf{x}}$ satisfies $\|\text{Conv}_{\overline{\mathbf{s}}}(\hat{\mathbf{x}})\|_F^2 = 0$ and $M_{\Omega}\hat{\mathbf{x}} = \mathbf{y}_{\Omega}$, the fixed point $\hat{\mathbf{x}}^{\infty}$ to which Algorithm (\ref{eq:7}) converges must be located within the following subspace:
\begin{equation}
\begin{aligned}
    \mathbb{U} :&= \{ \hat{\mathbf{x}} |  M_{\Omega}\hat{\mathbf{x}} = \mathbf{y}_{\Omega}, \|\text{Conv}_{\overline{\mathbf{s}}}(\hat{\mathbf{x}})\|_F^2 = 0 \}\\
    &\subseteq \{ \hat{\mathbf{x}} |  M_{\Omega}\hat{\mathbf{x}} = \mathbf{y}_{\Omega}, \text{rank}(\mathcal{H}(\hat{\mathbf{x}}))\leq r \}
\end{aligned} 
\end{equation} 
\begin{thm}[\cite{7744614}, informal]\label{thm:2} Under certain conditions, $\mathbb{U}=\{\hat{\mathbf{x}}^\dag \}$, where $\hat{\mathbf{x}}^\dag $ is the unique solution of $M_{\Omega}\hat{\mathbf{x}}^\dag = \mathbf{y}_{\Omega}$ with  $\text{rank}(\mathcal{H}(\hat{\mathbf{x}}^\dag))=r$.
\end{thm}
The above theorem illustrates that once the network (\ref{eq:7}) can converge to a fixed point, it can completely reconstruct the missing $k$-space data.
\subsection{Self-Supervised Learning}
To ensure the convergence of (\ref{eq:7}), we introduce the DEQ strategy. Before training the network (\ref{eq:7}) for parameter backpropagation, we first ensure that (\ref{eq:7}) can converge to a fixed point. Specifically, 
$\mathcal{F}_{\mathbf{s}}:= \mathcal{P}(\mathcal{G}_{\mathbf{s}}(\cdot))$ is an abbreviation of algorithm \eqref{eq:7}, then its fixed point is $ \hat{\mathbf{x}}^{\infty} = \mathcal{F}_{\mathbf{s}}(\hat{\mathbf{x}}^{\infty})$, where $ \hat{\mathbf{x}}^{\infty} := \hat{\mathbf{x}}^{\infty}(\mathbf{y}_{\Omega}, \mathbf{s})$. 
According to the chain rule, the partial derivative of loss function $ \ell (\hat{\mathbf{x}}^{\infty}(\mathbf{y}, \mathbf{s}), \cdot) $ at $\mathbf{s} $ is
\begin{equation}
   \frac{\partial \ell (\hat{\mathbf{x}}^{\infty}(\mathbf{y}_{\Omega}, \mathbf{s}), \cdot) }{\partial \mathbf{s}} = \frac{\partial \hat{\mathbf{x}}^{\infty}(\mathbf{y}_{\Omega}, \mathbf{s})^{T} }{\partial \mathbf{s}} \cdot \frac{\partial \ell (\hat{\mathbf{x}}^{\infty}(\mathbf{y}_{\Omega}, \mathbf{s}), \cdot) }{\partial \hat{\mathbf{x}}^{\infty}(\mathbf{y}_{\Omega}, \mathbf{s})} 
\end{equation}
Since 
\begin{equation}
    \frac{\partial \hat{\mathbf{x}}^{\infty} }{\partial \mathbf{s}} = 
\frac{\partial \mathcal{F}_{\mathbf{s}}(\hat{\mathbf{x}}^{\infty}) }{\partial \mathbf{s}} + \frac{\partial \mathcal{F}_{\theta}(\hat{\mathbf{x}}^{\infty}) }{\partial \hat{\mathbf{x}}^{\infty}} \cdot \frac{\partial \hat{\mathbf{x}}^{\infty} }{\partial \mathbf{s} }
\end{equation}
then
\begin{equation}
    \frac{\partial \hat{\mathbf{x}}^{\infty} }{\partial \mathbf{s} } = \left(\mathcal{I} - \frac{\partial \mathcal{F}_{\mathbf{s}}(\hat{\mathbf{x}}^{\infty}) }{\partial \hat{\mathbf{x}}^{\infty}}\right)^{-1} \frac{\partial \mathcal{F}_{\mathbf{s}}(\hat{\mathbf{x}}^{\infty}) }{\partial \mathbf{s} }
\end{equation}
Combining the above equations, we can obtain
\begin{equation}\label{eq:bp}
\frac{\partial \ell (\hat{\mathbf{x}}^{\infty},\cdot) }{\partial \mathbf{s}} =  \frac{\partial \mathcal{F}_{\mathbf{s}}(\hat{\mathbf{x}}^{\infty})^{T} }{\partial \mathbf{s}}  \left(\mathcal{I} - \frac{\partial \mathcal{F}_{\mathbf{s}}(\hat{\mathbf{x}}^{\infty}) }{\partial \hat{\mathbf{x}}^{\infty}}\right)^{-T} \frac{\partial \ell (\hat{\mathbf{x}}^{\infty}, \cdot) }{\partial \hat{\mathbf{x}}^{\infty}} 
\end{equation} 
From the above formula, it can be found that it only calculates the partial differential in $\hat{\mathbf{x}}^{\infty}$ and has nothing to do with the iterates of (\ref{eq:7}). It means that 
the backpropagation for kernel $\mathbf{s}$ can be calculated on fixed point $\hat{\mathbf{x}}^{\infty}$ directly, regardless of how many iterations are carried out, so that the memory does not increase even if the number of layers increases to infinity. 

\begin{figure*}[!t]
\centering
\includegraphics[width=0.95\textwidth,height=0.37\textwidth]{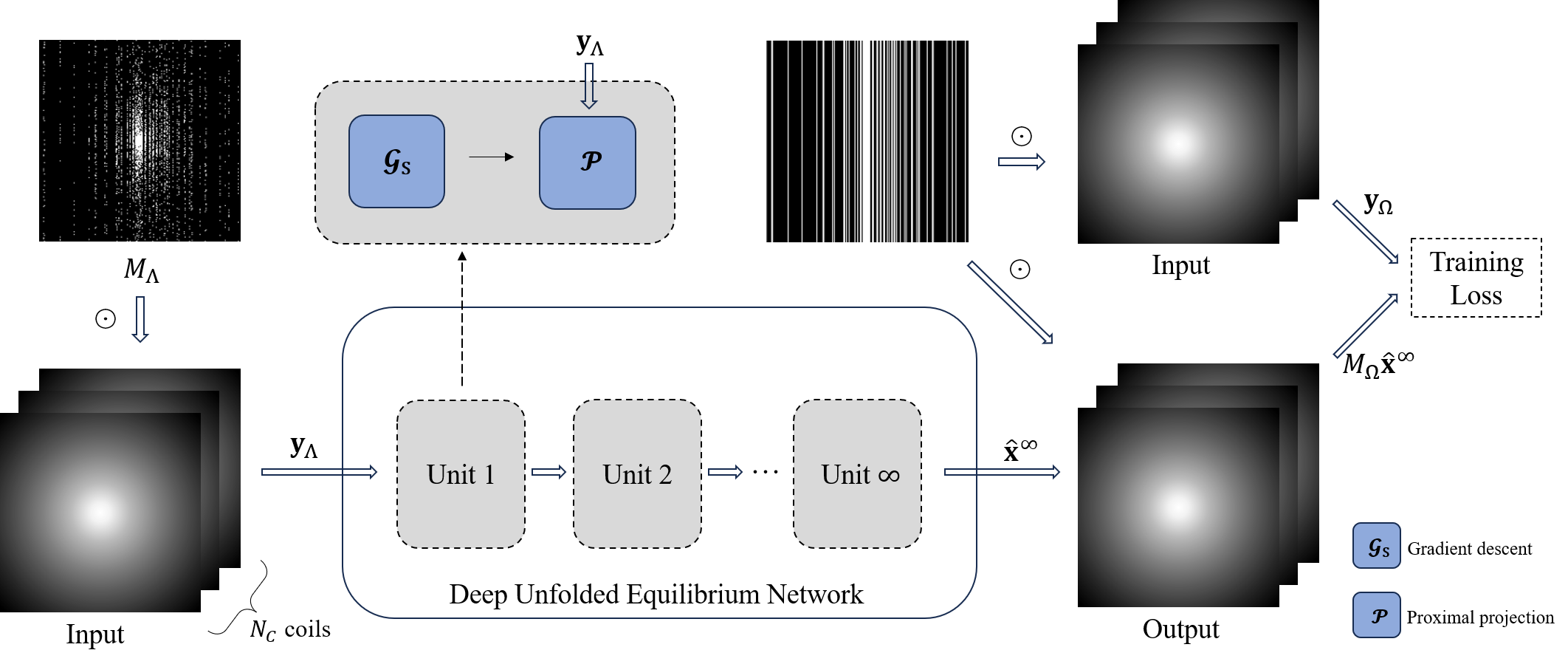}
\caption{Self-supervised deep PGD unfolded equilibrium network schematic. The binary sub-mask $M_\Lambda$ is indexed by the subset $\Lambda$ of the sampling location set $\Omega$. The point-wise product $\odot$ represents the under-sampling process. The sub-sampled $k$-space data $\mathbf{y}_\Lambda$ serves as input to the training network, while the available measurements $\mathbf{y}_\Omega$ are utilized to define the loss function. The network's outputs consist of multi-channel data in $k$-space, which are then compared with the corresponding measurements $\mathbf{y}_\Omega$.}
\label{Fig1}
\end{figure*}

In general supervised learning, the loss function takes the form:
\begin{equation}\label{eq:8}
 \min_{\theta} \frac{1}{N}\sum_{i=1}^{N}\ell(\hat{\mathbf{x}}^{\infty}(\mathbf{y}_{\Omega}^i, \mathbf{s}), \hat{\mathbf{x}}_{i}^{\text{ref}})
\end{equation} 
where $N$ is the number of elements in the training database, $\hat{\mathbf{x}}_{i}^{\text{ref}}$ is the $i$th fully sampled label in the dataset and $\mathbf{y}_{\Omega}^i$ is its corresponding undersampled $k$-space data. 
As previously mentioned, many factors can make it difficult or impossible to obtain fully sampled data. The practicality of DL based reconstruction methods that rely on supervised learning faces an important challenge due to the absence of reliable ground truth data for training. 
This shortcoming is overcome by the emergence of self-supervision \cite{yaman2020self}, which divides the sampled measurement $k$-space locations into two sets as follows  
\begin{equation}
     \Omega = \Gamma + \Lambda
\end{equation}
and takes one of them for training.
In our self-supervised strategy, the two sets of $k$-space points are not disjoint $ \Gamma \cap \Lambda = \phi $. Subsubsampling data $\mathbf{y}_{\Lambda} = M_{\Lambda}\mathbf{y}_{\Omega}$, let $(\mathbf{y}_{\Lambda},\mathbf{y}_{\Omega})$ as a training paired data to define the loss function. Specifically, the self-supervised training loss function takes the form
\begin{equation}\label{eq:9}
 \min_{\theta} \frac{1}{N}\sum_{i=1}^{N}\ell(M_{\Omega}\hat{\mathbf{x}}^{\infty}(\mathbf{y}_{\Lambda}^i, \mathbf{s}), \mathbf{y}_{\Omega}^i)
\end{equation} 
Specifically, we illustrate the self-supervised approach in Figure \ref{Fig1} and depict the training process in Algorithm \ref{alg: alg3}.

\begin{algorithm}[H]
\caption{Self-supervised Training.} \label{alg: alg3}
\begin{algorithmic}
\STATE 
\STATE {\textbf{ Input: }} training samples $\{(\mathbf{y}_{\Lambda}^i, \mathbf{y}_{\Omega}^i)\}_{i=1}^N$.
\STATE {\textbf{ Initialize:  }} $\mathbf{s}^{0}$.
\STATE  \hspace{0.25cm}{\textbf{for}} $p = 0,\dots, P-1$ \textbf{do}
\STATE \hspace{0.5cm} {\textbf{for}} $i =0,\dots, N-1$ \textbf{do}
\STATE \hspace{1cm} Carry out \eqref{eq:7} to find a fixed point:
\STATE \hspace{1.5cm}
    $\hat{\mathbf{x}}^{\infty}(\mathbf{y}_{\Lambda}^i, \mathbf{s}^{pN+i}) = \mathcal{P}(\mathcal{G}(\hat{\mathbf{x}}^{\infty}(\mathbf{y}_{\Lambda}^i, \mathbf{s}^{pN+i})))$
\STATE \hspace{1cm} $ \mathbf{s}^{pN+i+1} = \text{ ADAM }(\ell(M_{\Omega}\hat{\mathbf{x}}^{\infty}(\mathbf{y}_{\Lambda}^i, \mathbf{s}), \mathbf{y}_{\Omega}^i)) $ 
\STATE \hspace{0.5cm} {\textbf{end for}}
\STATE \hspace{0.25cm}{\textbf{end for}} 
\STATE {\textbf{ Output: }} $ \mathbf{s}^{PN}$
\end{algorithmic}
\end{algorithm}

The training is conducted over a total of $P$ epochs. In practice, to accelerate the convergence rate to find the fixed points, we employ Anderson acceleration \cite{Walker2011Anderson}. Additionally, for the convolutional kernel $\mathbf{s}$, we update it using the ADAM optimizer based on derivative rules (\ref{eq:bp}).

By inserting the well-trained convolutional kernel $ \mathbf{s}^{PN}$ into the network (\ref{eq:7}) and searching for its fixed points, we can achieve the reconstruction of missing $k$-space data. The specific process is demonstrated in Algorithm  \ref{alg: alg2}. 

\begin{algorithm}[H]
\caption{Testing.} \label{alg: alg2}
\begin{algorithmic}
\STATE 
\STATE {\textbf{ Input: }} testing samples $\{\mathbf{y}_{\Omega}^i\}_{i=1}^M$ and $\mathbf{s}^{PN}$ 
\STATE \hspace{0.5cm} {\textbf{for}} $i =0,\dots, M-1$ \textbf{do}
\STATE \hspace{1cm} Carry out \eqref{eq:7} to find a fixed point:
\STATE \hspace{1.5cm}
    $\hat{\mathbf{x}}^{\infty}(\mathbf{y}_{\Omega}^i, \mathbf{s}^{PN}) = \mathcal{P}(\mathcal{G}(\hat{\mathbf{x}}^{\infty}(\mathbf{y}_{\Omega}^i, \mathbf{s}^{PN})))$
\STATE \hspace{0.5cm} {\textbf{end for}}
\STATE {\textbf{ Output: }} $ \{\hat{\mathbf{x}}^{\infty}(\mathbf{y}_{\Omega}^i, \mathbf{s}^{PN})\}_{i=1}^M$.
\end{algorithmic}
\end{algorithm}

\subsection{Network Architecture Generalization} 
\label{subsec: framework}
\begin{figure}[t]
\centering  
\includegraphics[width=0.42\textwidth,height=0.6\textwidth]{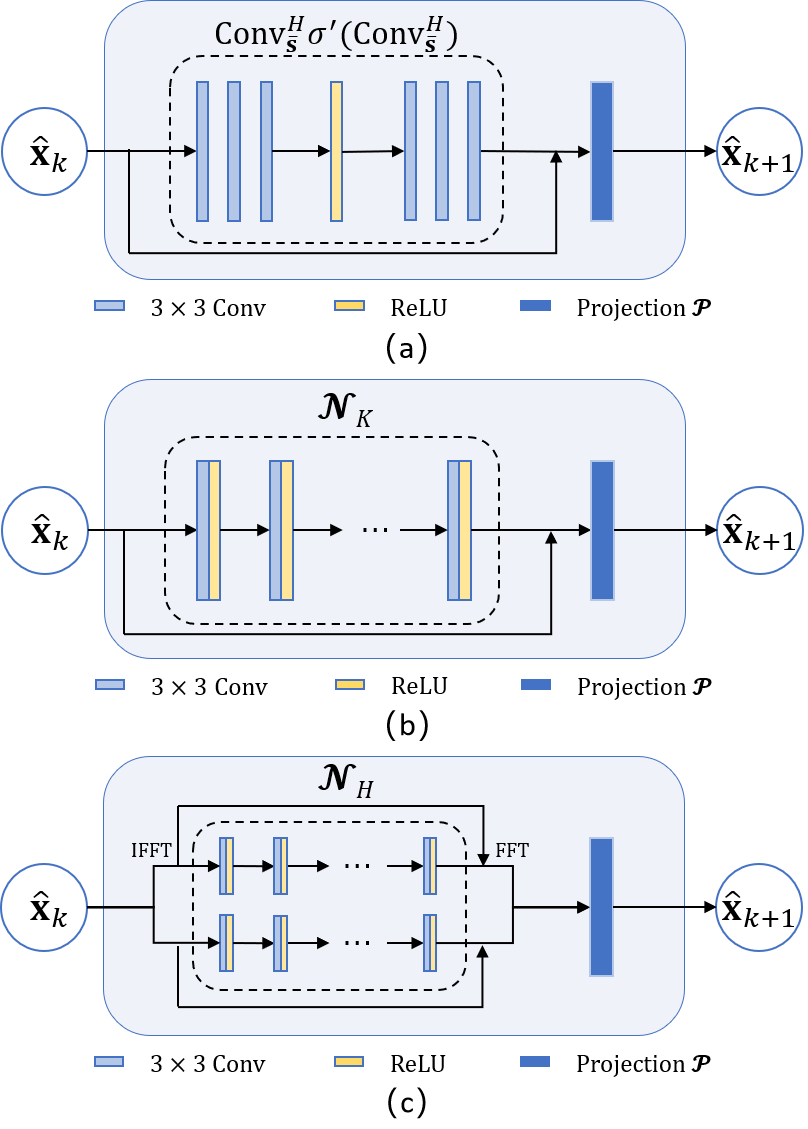}   
\caption{
Schematic diagram illustrating the architecture of each unit in the unfolded network. The primary distinction among the three gradient descent step schemes proposed in Section \ref{subsec: framework} lies in the design of the residual network. (a) employs convolutional and transpose convolutional layers to learn the null space filter of the SLR matrix. (b) is a generalized five-layer convolutional network in the $k$-space domain. (c) represents a hybrid network formed through a linear combination of the $k$-space domain network and an image domain network. The latter is also a five-layer convolutional network, assuming that the inverse Fourier transform has converted the signal into the image domain before returning to $k$-space.
}
\label{Fig: network}
\end{figure}

In network (\ref{eq:7}), at each iteration, the network module consists of a residual network composed of convolutions and transposed convolutions, without any activation functions. However, based on past experience, activation functions can play a crucial role. To address this, we generalize model (\ref{eq:slr2}) to the following form:

\begin{equation}\label{eq:slr3}
\min_{\hat{\mathbf{x}}} \sigma(\text{Conv}_{\overline{\mathbf{s}}}(\hat{\mathbf{x}}))~\text{s.t.}~M_{\Omega}\hat{\mathbf{x}} = \mathbf{y}_{\Omega}
\end{equation}
 where $\sigma$ is a convex positive penalty function that replaces the Frobenius norm. The corresponding PGD algorithm is as follows:
\begin{equation}\label{eq:7:2}
    \left\{\begin{aligned}
    \mathbf{r}^{k+1} &= \hat{\mathbf{x}}^{k} - \eta \text{Conv}_{\overline{\mathbf{s}}} ^{H}\sigma'(\text{Conv}_{\overline{\mathbf{s}}}(\hat{\mathbf{x}}^{k}))\\
    \hat{\mathbf{x}}^{k+1} &= \mathcal{P}_{\{ \hat{\mathbf{x}}|M_{\Omega}\hat{\mathbf{x}} = \mathbf{y}_{\Omega}\}} (\mathbf{r}^{k+1})
    \end{aligned}\right. 
\end{equation}
Here, $\sigma'$ acts as an activation function, and in practice, we chose $\sigma'=\text{ReLU}$. The schematic of the above self-supervised PGD unfolding network, termed SSPGD, is illustrated in Figure \ref{Fig: network} (a).

Practical experience has shown that improving the degree of parameterization and the capability of nonlinear representation of the network can effectively enhance the model. We generalize $\text{Conv}{\overline{\mathbf{s}}} ^{H}\sigma'(\text{Conv}{\overline{\mathbf{s}}}(\cdot))$ to a five-layer convolution function $\mathcal{N}_{K}(\cdot)$, which includes ReLU activation functions:
$\text{Conv}_{\overline{\mathbf{s}}} ^{H}\sigma'(\text{Conv}_{\overline{\mathbf{s}}}(\cdot))$ to a five-layer convolution function $ \mathcal{N}_{K}(\cdot)$ including ReLU activation functions, i.e., 
\begin{equation}\label{eq:11}
  \mathbf{r}^{k+1} = \hat{\mathbf{x}}^{k} - \eta \mathcal{N}_{K}(\hat{\mathbf{x}}^{k})
\end{equation} 
The architecture of the above $k$-space generalized self-supervised PGD unfolding network, termed KSSPGD, is described in Figure \ref{Fig: network} (b).

The two network architectures mentioned above focus on learning features in $k$-space. However, MR images typically have self-redundancies and complementary priors that need to be considered in the image domain. Following \cite{Pramanik2020DeepSLR}, we adopt a hybrid architecture. Specifically, the hybrid network $\mathcal{N}_{H}(\cdot)$ capitalizes on the annihilation relation in $k$-space and priors in the image domain simultaneously. The hybrid network $\mathcal{N}_{H}(\cdot)$ is defined as:
\begin{equation}\label{eq:12}
\begin{aligned}
  \mathbf{r}^{k+1} &= \hat{\mathbf{x}}^{k} - \lambda \mathcal{N}_{H}(\hat{\mathbf{x}}^{k}) \\
  &= \hat{\mathbf{x}}^{k} - \eta_{1} \mathcal{N}_{K}(\hat{\mathbf{x}}^{k}) - \eta_{2} \mathcal{N}_{I}(\hat{\mathbf{x}}^{k})
\end{aligned}
\end{equation} 
The architecture of the above hybrid $k$-space and image domain generalized self-supervised PGD unfolding network, termed HSSPGD, is described in Figure \ref{Fig: network} (c).

\section{Implementation}
\label{sec: implementation}
The evaluation was conducted on knee and brain MR data acquired by various $k$-space trajectories. The MR data and implementation details are elaborated below:

\subsection{Data Acquisition}
\subsubsection{Knee data}
The knee raw data \footnote{\url{https://fastmri.org/}} was acquired from a 3T Siemens scanner (Siemens Magnetom Skyra, Prisma and Biograph mMR). Data acquisition used a 15 channel knee coil array and conventional Cartesian 2D TSE protocol employed clinically at NYU School of Medicine. The following sequence parameters were used: Echo train length 4, matrix size $368 \times 368$, in-plane resolution $0.5mm\times0.5mm$, slice thickness $3mm$, no gap between slices. Timing varied between systems, with repetition time (TR) ranging between 2200 and 3000 milliseconds, and echo time (TE) between 27 and 34 milliseconds. We randomly selected 31 individuals (840 slices in total) as training data and 3 individuals (96 slices in total) as test data.

\subsubsection{Brain data}
The brain data used for this study was acquired from \cite{Aggarwal2019MoDL}. Data acquisition used a 3D T2 CUBE sequence with Cartesian readouts and a 12-channel head coil array. 
Data were fully sampled from five subjects, each yielding matrices dimensions were $256 \times 232 \times 208$ with $1 mm$ isotropic resolution. For the initial four subjects, a selection of 90 slices with parts of the anatomy was made from a total of 208 slices for training. 120 slices with clear and effective images from the fifth subject were reserved for testing. We conducted retrospective undersampling for training and evaluating. All experiments about brain data utilize a variable-density Cartesian random sampling mask with different undersampling factors. The coil sensitivity maps for comparative trial were estimated from the central $k$-space regions of each slice according to ESPIRiT \cite{Uecker2014ESPIRiT}.

\subsubsection{Sampling trajectories}
\begin{figure}[!t]
\centering  
{\label{Fig.mask1}
\includegraphics[width=0.1\textwidth,height=0.1\textwidth]{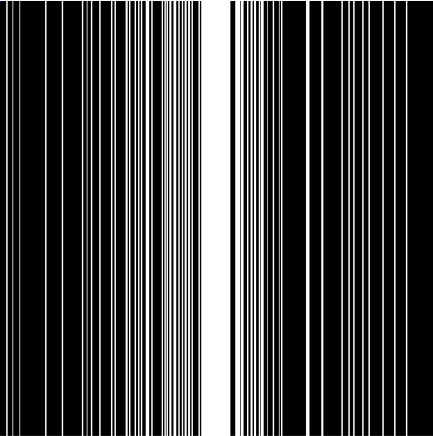} }
{\label{Fig.mask2}
\includegraphics[width=0.1\textwidth,height=0.1\textwidth]{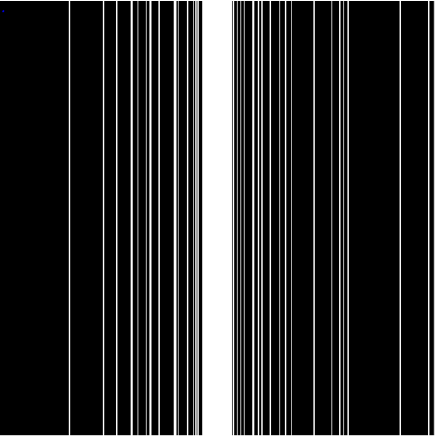}  }
{\label{Fig.mask3}
\includegraphics[width=0.09\textwidth,height=0.1\textwidth]{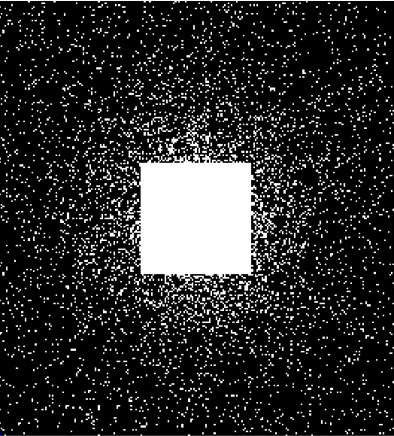}   }
{\label{Fig.mask4}
\includegraphics[width=0.09\textwidth,height=0.1\textwidth]{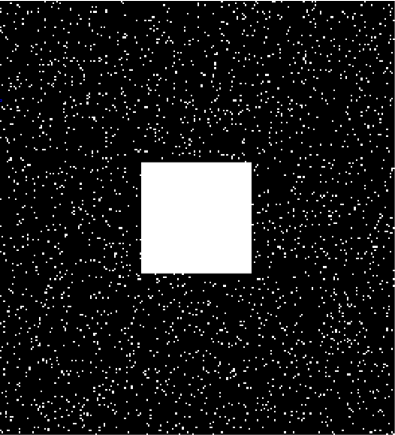}   }
\caption{ Various sampling trajectories: (a) 1D random undersampling at $R=4$, (b) 1D regular undersampling at $R=6$, (c) 2D regular undersampling at $R=6$, and (d) 2D regular undersampling at $R=10$.
}
\label{Fig: mask}
\end{figure}

We apply different undersampling trajectories in comparison experiments to demonstrate the trustworthiness of our model.  Fig.~\ref{Fig: mask} illustrates a visualization of undersampling trajectories.  For 2D dataset,   the corresponding sampling trajectories are randomly distributed with $\text{ACS}=24$. Proposed reconstruction runs at acceleration $R = 4, 6$. The performance of our method corresponding for 3D dataset is randomly undersampled at different acceleration $R=6,10$  with $\text{ACS} = 48\times 48$. Sub-undersampling for self-supervision accounts for about 50\% of undersampling. 
\\
\\
\\

\subsection{Network Architecture and Training}

The architecture schematic diagram of our proposed mddel is illustrated in Fig.~\ref{Fig: network}. Particularly, given the complex nature of the data, the real and imaginary components were concatenated before being fed into the network. We use $3\times3$ convolution kernel with 64 channels to achieve convolution and the layers of our unfolding network is set to 10.
Optimizer is ADAM \cite{kingma2014adam} with $\beta_1=0.9, \beta_2=0.999$ for optimizing loss function \eqref{eq:9}. The learning rate is $10^{-4}$. 
The experiments were implemented using an Ubuntu 20.04 operating system and a NVIDIA A100 Tensor Core GPU with 80 GB memory. The open-source PyTorch 1.10 framework \cite{paszke2019pytorch} was utilized, along with CUDA 11.3 and CUDNN support for efficient computation.

\subsection{Performance Evaluation}

In this study, to avoid potential bias introduced by varying image merging methods during the quantification of differently reconstructed images, both the reconstructed and reference images underwent an inverse Fourier transform followed by ESPIRiT-estimated sensitivity map merging. The quantitative evaluations was calculated within the image domain. Quantitative evaluation adopted the peak signal-to-noise ratio (PSNR), normalized mean square error (NMSE) value, and the structural similarity (SSIM) index \cite{Wang2004SSIM} for reference images and parallel reconstructions using various methods.

\section{Experimentation Results}
\label{sec: experimentation}

\subsection{Ablation Studies}

\begin{figure}[!t]
\centering
\includegraphics[width=0.47\textwidth,height=0.47\textwidth]{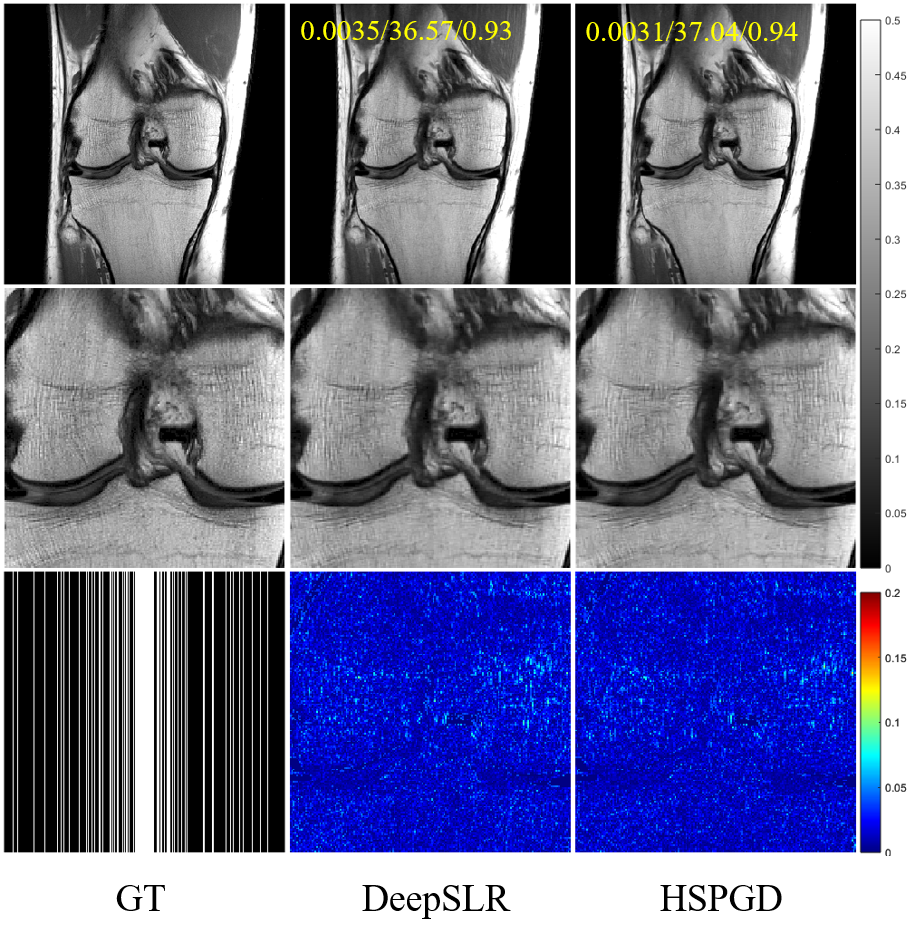}
\caption{
Reconstructions of supervised models at random undersampling $R=4$ with 24 ACS lines. The leftmost column shows the ground truth and undersampling mask. The data on top represents the NMSE/PSNR/SSIM values of this slice. The second row displays magnified details. The third row presents the error maps of magnified versions. The grayscale of reconstructed images and color bars of error maps are located on the right side.
}
\label{Fig: super}
\end{figure}

First, we eliminate the influence of self-supervised learning and compare the DeepSLR with our method in a supervised learning scenario to verify the advantages of a rigorously model-driven network structure.
For a fair comparison, we select the supervised version of HSSPGD, termed HSPGD, as the comparative method, which has network parameters and results most consistent with DeepSLR. The reconstructed results are shown in Fig. \ref{Fig: super}. From the error view, it is not hard to find that HSPGD achieves more accurate reconstructions. The quantitative metrics are shown in Table \ref{tab: super} which further validates the superiority of the proposed HSPGD. Therefore, the above experiments confirmed the effectiveness of the proposed rigorously model-driven network structure.

\begin{table}[!t]
\caption{Comparison of quantitative experimental results for supervised studies on knee dataset using random undersampling, with the optimal values highlighted in \textcolor{red}{red}.}
\centering
  \footnotesize
      \begin{tabular}{c|l|cccc}
        \hline \hline 
	\multicolumn{ 2}{c}{ Undersampling} & \multicolumn{ 3}{|c}{Quantitative Evaluation}  \\
	\multicolumn{ 2}{c|}{ \& Methods   } &NMSE &PSNR(dB)&SSIM   \\ 
        \hline  
				\multirow{2}{*}{\makecell{2D \\ ($R=4$)}}
				&  DeepSLR & 0.0047$\pm$0.0036 & 36.03$\pm$2.05 & 0.90$\pm$0.12 \\				
				\cline{2-5}
                &  HSPGD & \textcolor{red}{0.0044$\pm$0.0036} & \textcolor{red}{36.34$\pm$2.12} & \textcolor{red}{0.91$\pm$0.10} \\
        \hline \hline
    \end{tabular}
\label{tab: super}
\end{table}

 \subsection{Comparative Studies}
\begin{figure}[!t]
\centering
\includegraphics[width=0.49\textwidth,height=0.38\textwidth]{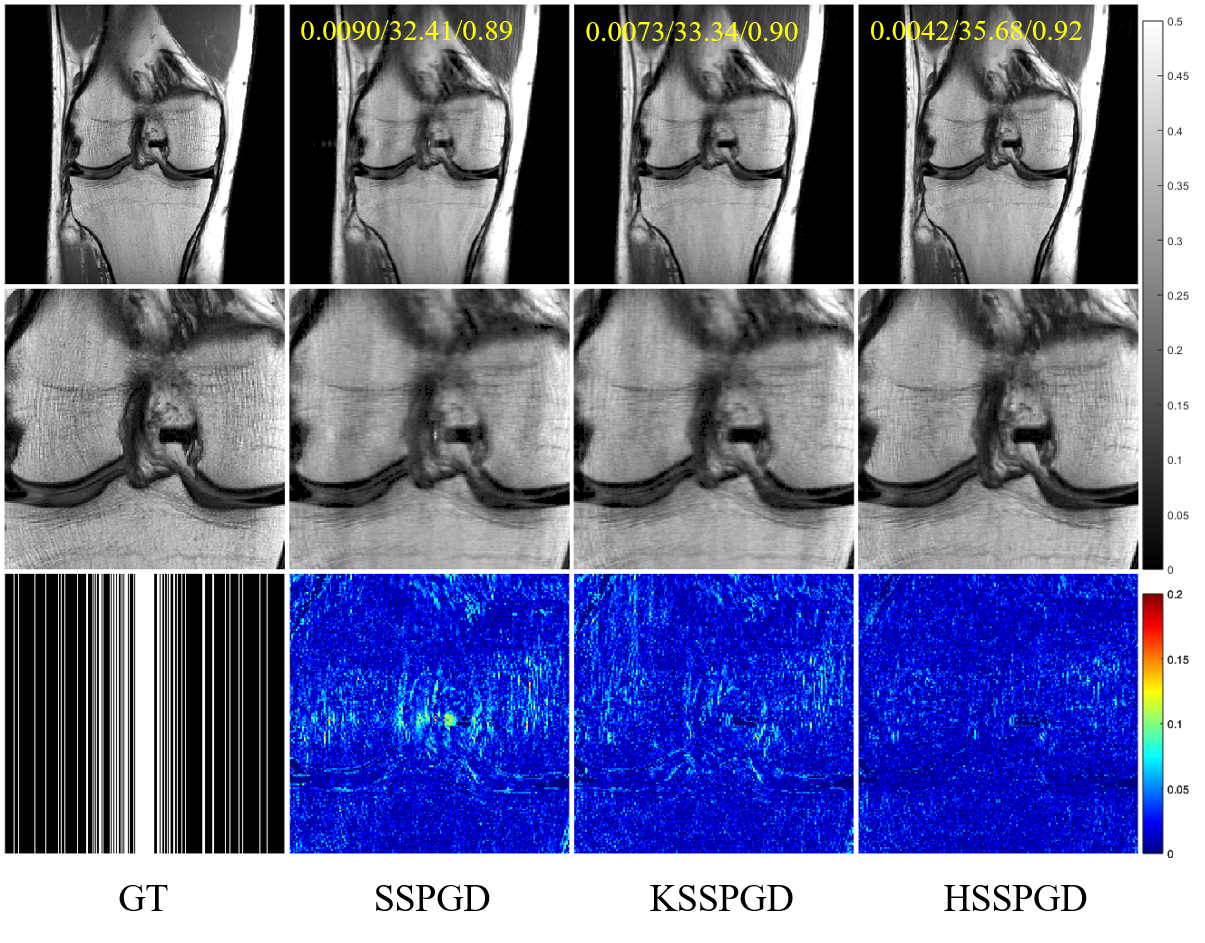}
\caption{
Reconstructions of different models at random undersampling $R=4$ with 24 ACS lines. The leftmost column shows the ground truth and undersampling mask. The data on top represents the NMSE/PSNR/SSIM values of this slice. The second row displays magnified details. The third row presents the error maps of magnified versions. The grayscale of reconstructed images and color bars of error maps are located on the right side.
}
\label{Fig: ablation}
\end{figure}

Before comparing our method with existing approaches, we conducted an internal performance evaluation of the three network architectures proposed in this paper, namely, SSPGD, as well as the generalized KSSPGD and HSSPGD. Specifically, we implemented the SSPGD model in the $k$-space domain using convolutional and transpose convolutional layers to learn the null space filter $\text{Conv}_{\overline{\mathbf{s}}}$. We implemented the KSSPGD model in the $k$-space domain with 5 convolutional layers activated by the ReLU function. Lastly, we implemented the HSSPGD model in both the $k$-space and image domains with a hybrid structure comprising 5 convolutional layers activated by the ReLU function. Fig.~\ref{Fig: ablation} displays the reconstruction results of our proposed method with three different network architectures. Among them, HSSPGD, which leverages both $k$-space priors and image domain redundancy, exhibits the best performance. Table Fig.~\ref{Fig: ablation} presents the quantitative metrics, which also indicate that HSSPGD achieves the best results. Consequently, we will use HSSPGD, which demonstrates superior performance, as the comparative method for comparisons with existing approaches.

\begin{table}[!t]
\caption{Comparison of quantitative experimental results for ablation studies on knee dataset using random undersampling, with the optimal values highlighted in \textcolor{red}{red}.}
\centering
  \footnotesize
      \begin{tabular}{c|l|cccc}
        \hline \hline 
	\multicolumn{ 2}{c}{ Undersampling} & \multicolumn{ 3}{|c}{Quantitative Evaluation}  \\
	\multicolumn{ 2}{c|}{ \& Methods   } &NMSE &PSNR(dB)&SSIM   \\ 
        \hline  
				\multirow{3}{*}{\makecell{2D \\ ($R=4$)}}
				&  SSPGD & 0.0164$\pm$0.0122 & 30.69$\pm$2.74 & 0.85$\pm$0.13 \\
				\cline{2-5}
    		  &  KSSPGD  & 0.0120$\pm$0.0067 & 31.86$\pm$2.54 & 0.86$\pm$0.12 \\
				\cline{2-5}
				&  HSSPGD & \textcolor{red}{0.0055$\pm$0.0037} & \textcolor{red}{35.27$\pm$1.68} & \textcolor{red}{0.89$\pm$0.12} \\
        \hline \hline
    \end{tabular}
\label{tab: ablation}
\end{table}

Next, we compared our proposed model with various MRI reconstruction algorithms in different conditions to validate the effectiveness of our proposed model. These calibration-less and calibrated algorithms we compared against include a traditional reconstruction technique SPIRIT \cite{lustig2010spirit}, the early-stage self-supervised approach LORAKI \cite{kim2019loraki}, the state-of-the-art self-supervised method SSDU \cite{yaman2020self}, and our theoretically grounded self-supervised solution. Furthermore, we compare our self-supervised algorithm with the supervised algorithm DeepSLR \cite{Pramanik2020DeepSLR}, in order to clarify its gap with the supervised algorithms. Ensuring equitable environment, identical undersampling mask is used during training across all participating models in our comparative study. The evaluation of metrics was performed on background-removed images to ensure unbiased assessment.

\begin{figure*}[!t]
\centering
\includegraphics[width=1\textwidth,height=0.54\textwidth]{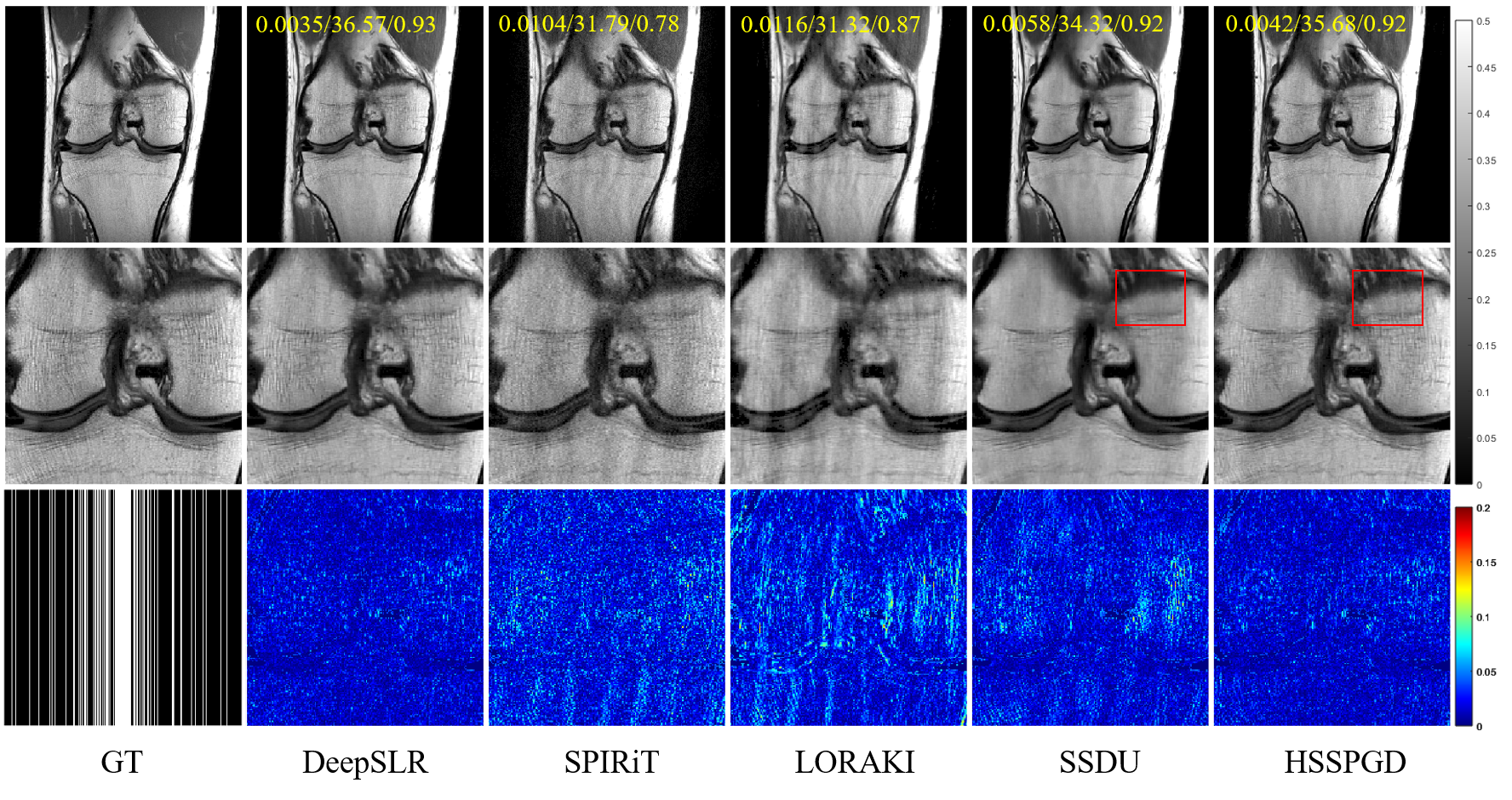}
\caption{
Reconstruction results under random undersampling at $R=4$ with 24 ACS lines. The leftmost column shows the ground truth and undersampling mask. The data on top represents the NMSE/PSNR/SSIM values of this slice. The second row displays magnified details. The third row presents the error maps of magnified versions. The grayscale of reconstructed images and color bars of error maps are located on the right side.
}
\label{knee4X24acs}
\end{figure*}

Firstly, we completed reconstruction task on knee data using various methods with random sampling. Figure \ref{knee4X24acs} illustrates the results of random sampling with an acceleration factor of 4 and 24 ACS lines. These outcomes of self-supervised reconstruction  exhibit varying degrees of aliasing artifacts. LORAKI achieves self-supervision through training the model using ACS lines solely, which result in more artifacts. In terms of details, SSDU demonstrates commendable reconstruction performance but generates blurry MR images lacking texture details (as indicated by the red box). Specially, HSSPGD is almost as effective as supervised algorithms and successfully suppresses aliasing artifacts to achieve satisfactory results. The mean values for different evaluation metrics on test set are listed in Table \ref{tab: knee}. Obviously, visually and in terms of error metrics, the reconstructed images of our method achieve optimal performance compared to other self-supervised algorithms. These qualitative and quantitative experimental results substantiate rationality of our previous analysis based on theoretical guarantees.

\begin{figure*}[!t]
\centering
\includegraphics[width=1\textwidth,height=0.54\textwidth]{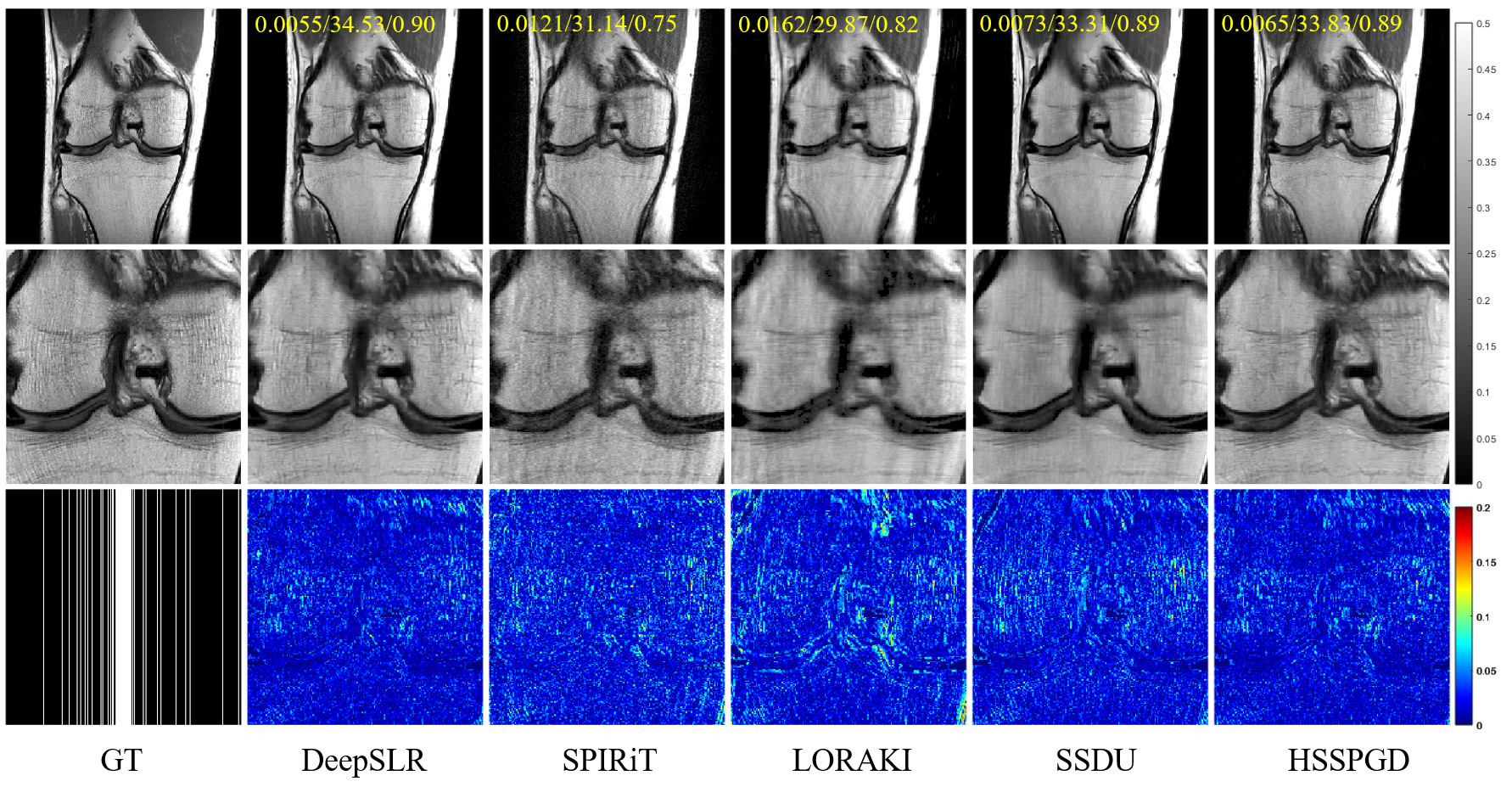}
\caption{Reconstruction results under random undersampling at $R=6$ with 24 ACS lines. The leftmost column shows the ground truth and undersampling mask. The data on top represents the NMSE/PSNR/SSIM values of this slice. The second row displays magnified details. The third row presents the error maps of magnified versions. The grayscale of reconstructed images and color bars of error maps are located on the right side. }
\label{knee6X24acs}
\end{figure*}

The comparison of various methods for knee data reconstruction under random sampling with an acceleration factor of 6 and 24 ACS lines are illustrated in Fig.~\ref{knee6X24acs}. Table \ref{tab: knee} presents the competitive quantitative results of these methods. With increasing acceleration factors, all methods exhibit higher errors; however, HSSPGD stands out as a self-supervised algorithm that closely approaches the reconstruction performance achieved by supervised learning.

\begin{figure*}[!t]
\centering
\includegraphics[width=0.99\textwidth,height=0.56\textwidth]{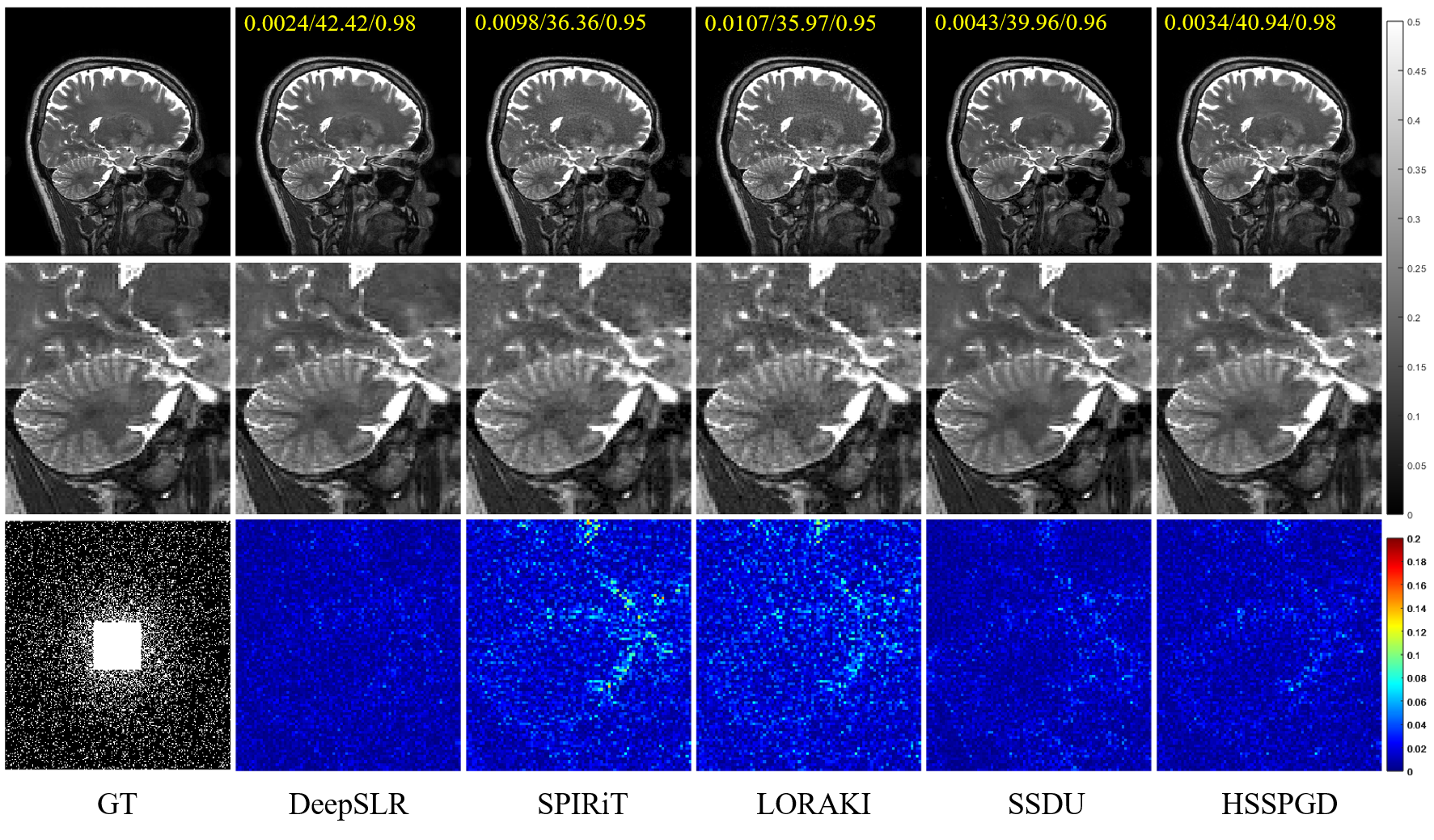}
\caption{Reconstruction results under random undersampling at $R=6$ with 48$\times$48 ACS regions. The leftmost column shows the ground truth and undersampling mask. The data on top represents the NMSE/PSNR/SSIM values of this slice. The second row displays magnified details. The third row presents the error maps of magnified versions. The grayscale of reconstructed images and color bars of error maps are located on the right side. }
\label{brain6X48acs}
\end{figure*}

\begin{figure*}[!t]
\centering
\includegraphics[width=0.99\textwidth,height=0.56\textwidth]{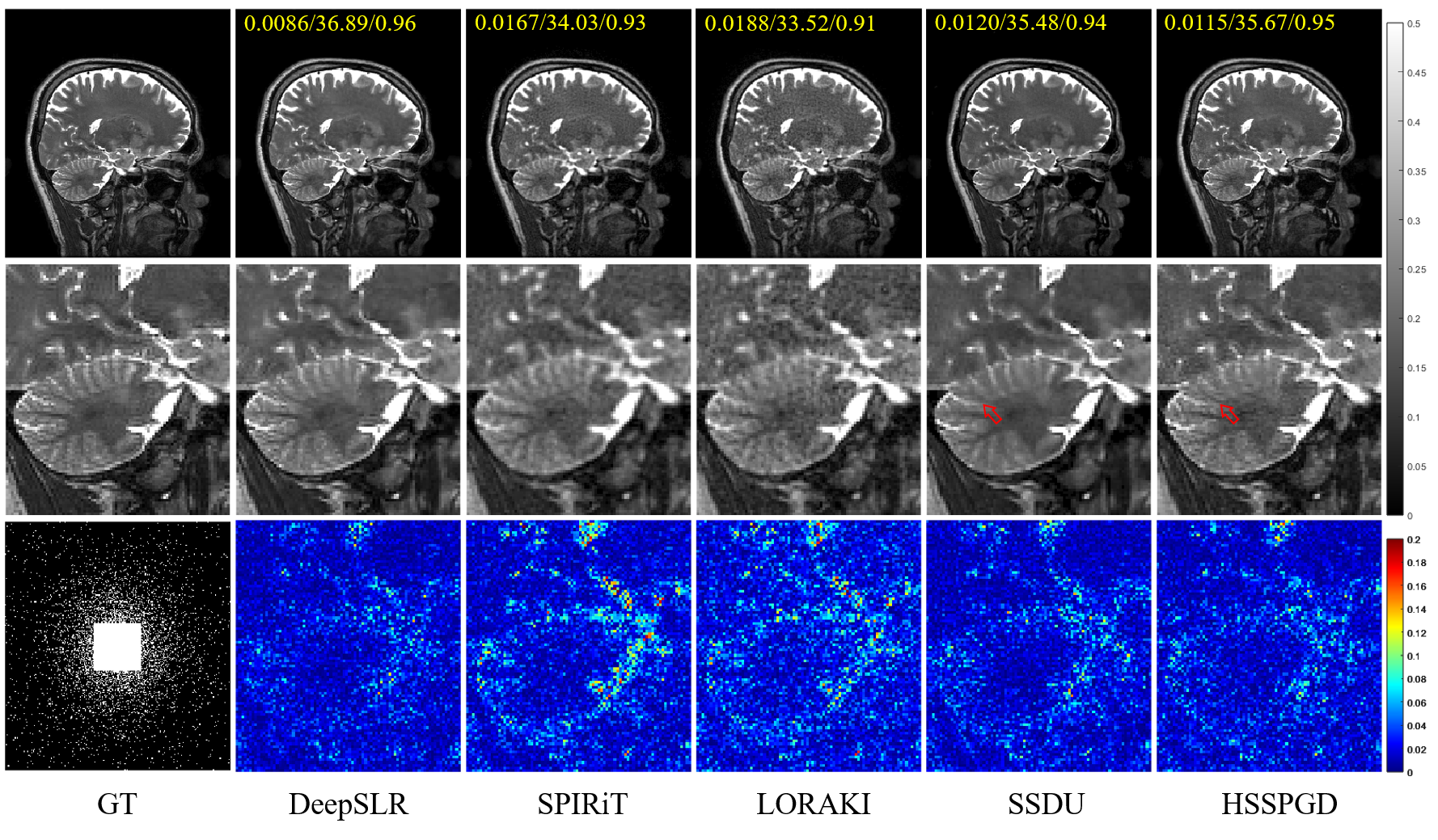}
\caption{Reconstruction results under random undersampling at $R=10$ with 48$\times$48 ACS regions. The leftmost column shows the ground truth and undersampling mask. The data on top represents the NMSE/PSNR/SSIM values of this slice. The second row displays magnified details. The third row presents the error maps of magnified versions. The grayscale of reconstructed images and color bars of error maps are located on the right side. }
\label{brain10X48acs}
\end{figure*}

In order to provide a general and adequate confirmation of the validity of our deep unfolded equilibrium models, we have also supplemented the reconstructed results of brain data by various methods. Fig.~\ref{brain6X48acs} and Fig.~\ref{brain10X48acs} respectively display the reconstruction of brain data with acceleration factors of 6 and 10 using random sampling. The quantitative indicators for the test dataset are presented in Table \ref{tab: brain}. Our model has a clear advantage at a 6-fold acceleration, achieving reconstructions closest to the supervised approach even in small brain regions with rich details. 
The location information provided by the sensitivity map plays a greater role as the acceleration is increased, so the SSDU works well at an acceleration factor of 10, but the organizational changes indicated by the arrow are still not recovered, which is mitigated on the HSSPGD.
In fact, reconstruction performance of our model is affected due to less sampling in high-frequency regions when accelerating at a larger factor because ACS occupies a significant proportion.

\begin{table}[!t]
\caption{Comparison of quantitative experimental results for various methods on knee dataset using random undersampling. Among all methods, the optimal values are denoted in \textcolor{red}{red}, while the optimal values for the self-supervised approach are highlighted in \textbf{bold}. }
\centering
  \footnotesize
      \begin{tabular}{c|l|cccc}
        \hline \hline 
	\multicolumn{ 2}{c}{ Undersampling} & \multicolumn{ 3}{|c}{Quantitative Evaluation}  \\
	\multicolumn{ 2}{c|}{ \& Methods   } &NMSE &PSNR(dB)&SSIM   \\ 
        \hline  
				\multirow{5}{*}{\makecell{Random \\($R=4$)}   }
     			    & DeepSLR & \textcolor{red}{0.0047$\pm$0.0036} & \textcolor{red}{36.03$\pm$2.05} & \textcolor{red}{0.90$\pm$0.12}   \\
                        \cline{2-5}
					& Spirit & 0.0171$\pm$0.0122 & 30.64$\pm$3.15 & 0.71$\pm$0.13  \\
					\cline{2-5}
					& Loraki  & 0.0176$\pm$0.0092 & 30.11$\pm$2.10 & 0.84$\pm$0.12  \\
					\cline{2-5}
					& SSDU & 0.0075$\pm$0.0063 & 34.08$\pm$2.42 & 0.88$\pm$0.13  \\
					\cline{2-5}
					& HSSPGD & \textbf{0.0055$\pm$0.0037} & \textbf{35.27$\pm$1.68} & \textbf{0.89$\pm$0.12}  \\
					\hline
				\multirow{5}{*}{\makecell{Random \\($R=6$)}}	
     			    & DeepSLR & \textcolor{red}{0.0074$\pm$0.0054} & \textcolor{red}{34.03$\pm$2.14} & \textcolor{red}{0.87$\pm$0.13}   \\
                        \cline{2-5}
					& Spirit  & 0.0181$\pm$0.0107 & 30.11$\pm$2.65 & 0.71$\pm$0.14  \\
					\cline{2-5}
					& Loraki  & 0.0204$\pm$0.0126 & 29.47$\pm$2.02 & 0.81$\pm$0.12  \\
					\cline{2-5}
					& SSDU & 0.0091$\pm$0.0051 & 32.96$\pm$2.13 & 0.85$\pm$0.13  \\
					\cline{2-5}
					& HSSPGD & \textbf{0.0081$\pm$0.0041} & \textbf{33.38$\pm$1.60} & \textbf{0.86$\pm$0.11}   \\
        \hline \hline
    \end{tabular}
\label{tab: knee}
\end{table}

\begin{table}[!t]
\caption{Comparison of quantitative experimental results for various methods on brain dataset using random undersampling. Among all methods, the optimal values are denoted in \textcolor{red}{red}, while the optimal values for the self-supervised approach are highlighted in \textbf{bold}.}
\centering
  \footnotesize
      \begin{tabular}{c|l|cccc}
        \hline \hline 
	\multicolumn{ 2}{c}{ Undersampling} & \multicolumn{ 3}{|c}{Quantitative Evaluation}  \\
	\multicolumn{ 2}{c|}{ \& Methods   } &NMSE &PSNR(dB)&SSIM   \\ 
        \hline  
				\multirow{5}{*}{\makecell{Random \\($R=6$)}   }		
     			    & DeepSLR & \textcolor{red}{0.0033$\pm$0.0016} & \textcolor{red}{41.34$\pm$1.37} & \textcolor{red}{0.98$\pm$0.00} \\
                        \cline{2-5}
					& Spirit  & 0.0082$\pm$0.0025 & 37.18$\pm$1.50 & 0.96$\pm$0.01 \\
					\cline{2-5}
					& Loraki  & 0.0100$\pm$0.0029 & 36.25$\pm$0.91 & 0.95$\pm$0.01 \\
					\cline{2-5}
					& SSDU & 0.0046$\pm$0.0011 & 39.60$\pm$1.35 & 0.96$\pm$0.01 \\
					\cline{2-5}
					& HSSPGD & \textbf{0.0041$\pm$0.0021} & \textbf{40.32$\pm$1.35} & \textbf{0.98$\pm$0.00} \\
					\hline
				\multirow{5}{*}{\makecell{Random \\($R=10$)}}
     			    & DeepSLR & \textcolor{red}{0.0092$\pm$0.0026} & \textcolor{red}{36.64$\pm$0.91} & \textcolor{red}{0.96$\pm$0.00} \\
                        \cline{2-5}
					& Spirit  & 0.0204$\pm$0.0049 & 33.12$\pm$1.12 & 0.93$\pm$0.01\\
					\cline{2-5}
					& Loraki  & 0.0245$\pm$0.0068 & 32.35$\pm$0.89 & 0.90$\pm$0.01 \\
					\cline{2-5}
					& SSDU & 0.0144$\pm$0.0143 & 34.72$\pm$1.64 & 0.94$\pm$0.01\\
					\cline{2-5}
					& HSSPGD & \textbf{0.0140$\pm$0.0030} & \textbf{34.72$\pm$1.02} & \textbf{0.95$\pm$0.01} \\
        \hline \hline
    \end{tabular}
\label{tab: brain}
\end{table}

\section{Discussion}
\label{sec: discussion}

\begin{figure}[!t]
\centering
\includegraphics[width=0.47\textwidth,height=0.5\textwidth]{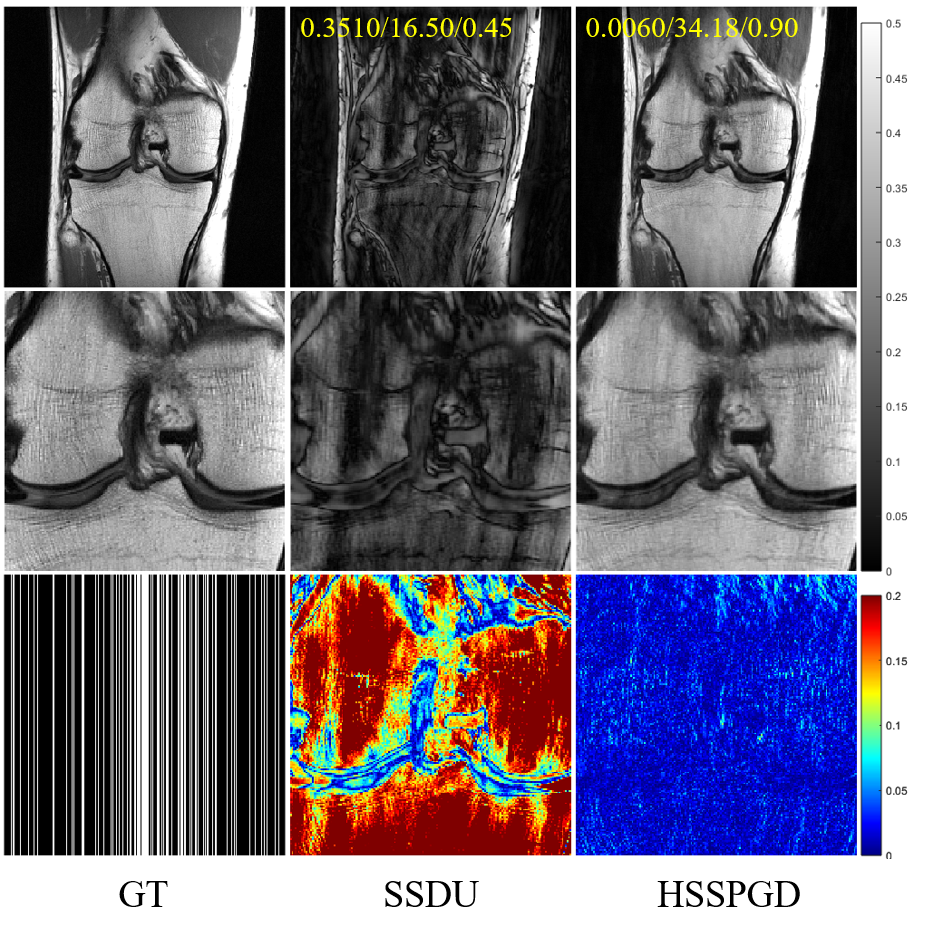}
\caption{Reconstruction results under random undersampling at $R=4$ with 8 ACS lines. The leftmost column shows the ground truth and undersampling mask. The data on top represents the NMSE/PSNR/SSIM values of this slice. The second row displays magnified details. The third row presents the error maps of magnified versions. The grayscale of reconstructed images and color bars of error maps are located on the right side. }
\label{Fig: kneecsm8}
\end{figure}

\begin{figure}[!t]
\centering
\includegraphics[width=0.47\textwidth,height=0.5\textwidth]{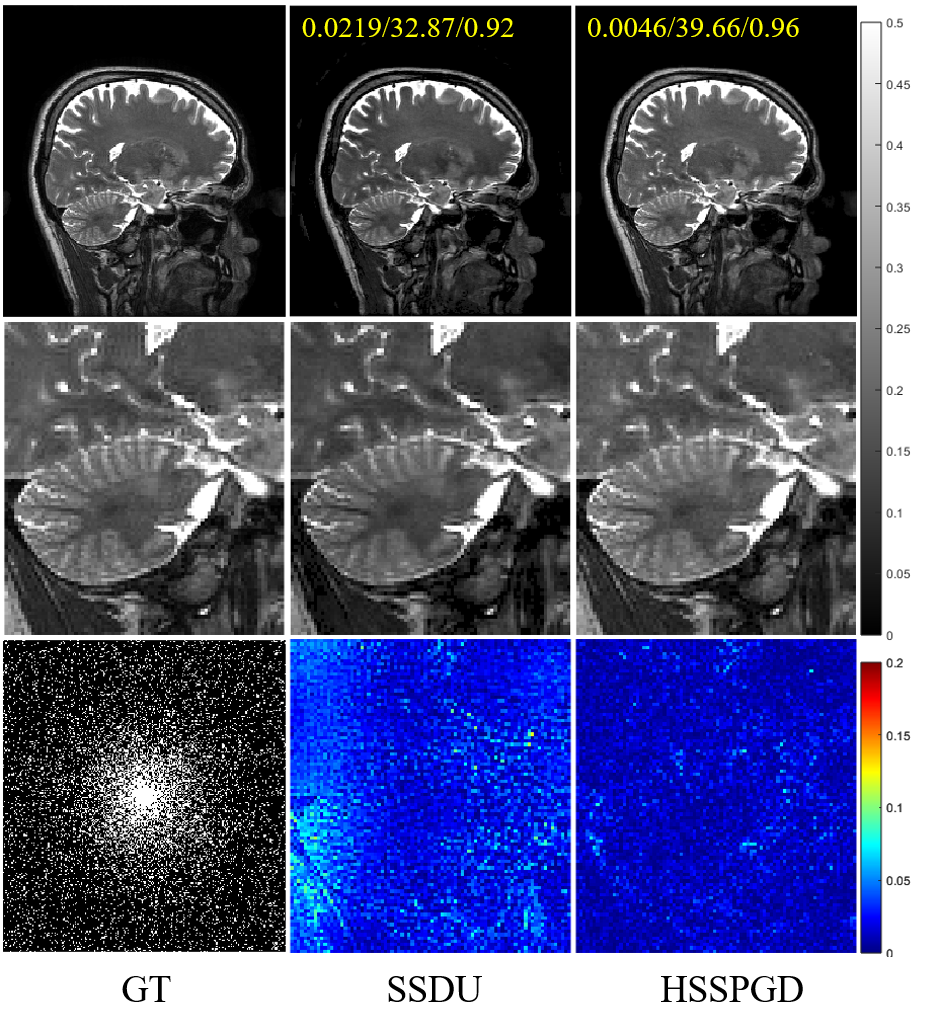}
\caption{Reconstruction results under random undersampling at $R=6$ with 12$\times$12 ACS regions. The leftmost column shows the ground truth and undersampling mask. The data on top represents the NMSE/PSNR/SSIM values of this slice. The second row displays magnified details. The third row presents the error maps of magnified versions. The grayscale of reconstructed images and color bars of error maps are located on the right side. }
\label{Fig: braincsm12}
\end{figure}

In the previous sections, this paper introduced a theoretically grounded $k$-space interpolation model suitable for self-supervised learning. Furthermore, the experiments demonstrate its superiority compared to other deep learning methods.

In these experiments, to ensure a fair comparison, the undersampled data contained a sufficient number of Auto-Calibrating Signal (ACS) lines to ensure accurate coil sensitivity estimation for image domain methods like SSDU. However, it's worth noting that this is unnecessary for the proposed methods. Therefore, we tested the performance of the proposed HSSPGD and SSDU when the number of ACS lines was limited. Fig. \ref{Fig: kneecsm8} and Fig. \ref{Fig: braincsm12} show the results of both approaches. Due to an insufficient number of ACS lines to support accurate coil sensitivity estimation, there was a significant deviation between the SSDU reconstructed images and the ground truth. In contrast, HSSPGD could still achieve accurate reconstruction. This demonstrates that the proposed $k$-space HSSPGD is more robust to changes in the undersampling patterns compared to image-domain reconstruction methods like SSDU.

On the other hand, the proposed method still has some limitations. Firstly, the proposed SSPGD network (\ref{eq:7:2}) is entirely consistent with $k$-space SLR model (\ref{eq:slr3}). However, as seen in Fig. \ref{Fig: ablation}, there is still a performance gap between SSPGD and its generalized versions, KSSPGD and HSSPGD. Therefore, improving the performance of SSPGD or providing full interpretability for KSSPGD and HSSPGD is part of our future work. Additionally, the theoretical guarantees (Theorems \ref{thm:1} and \ref{thm:2}) for the proposed method rely on learning the null space filter $ \mathbf{s}^{PN}$ accurately from the undersampled data, i.e., Algorithm \ref{alg: alg3}. While this is a common assumption in spatial SLR methods, there is currently no strict theoretical guarantee. Therefore, providing an error analysis for the $ \mathbf{s}^{PN}$ learned by Algorithm \ref{alg: alg3} is also part of our future work.

\section{Conclusion}
\label{sec: conclusion}
In this paper, we introduced an MC-informed deep unfolding equilibrium algorithm for $k$-space interpolation, eliminating the dependency on fully sampled data for supervision. Moreover, the MC theory's assurance ensures that our method essentially learns the principles of missing $k$-space data complete reconstruction from undersampled $k$-space data, establishing it as a ``white-box" approach. Furthermore, we provided proof of the convergence of the proposed unfolding network. Finally, our experiments not only validated the effectiveness of our approach but also demonstrated its superiority over existing self-supervised methods and traditional regularization techniques. In specific scenarios, our method achieved performance comparable to that of supervised learning methods.

\appendix
\subsection{Proof of Theorem \ref{thm:1}}\label{app:1}

\begin{proof}
Since  $\lambda_{\max}( \text{Conv}_{\overline{\mathbf{s}}} ^{H}\text{Conv}_{\overline{\mathbf{s}}}) \leq 1-\epsilon$ with $\epsilon<1$ and $\eta<1/(1-\epsilon)$, we have
\begin{equation} \label{eq: app1}
    \begin{aligned}
    \|\partial_{\hat{\mathbf{x}}} \mathcal{P}(\mathcal{G}_{\mathbf{s}}(\hat{\mathbf{x}}) ) \| 
    &= \| (I - M_{\Omega}) \partial_{\hat{\mathbf{x}}} \mathcal{G}_{\mathbf{s}}(\hat{\mathbf{x}})  \| \\
    &\le \| I- \eta \partial_{\hat{\mathbf{x}}}\text{Conv}_{\overline{\mathbf{s}}} ^{H}\text{Conv}_{\overline{\mathbf{s}}} (\hat{\mathbf{x}})  \| \\
    &\le (1-\eta) + \eta \| I -  \partial_{\hat{\mathbf{x}}}\text{Conv}_{\overline{\mathbf{s}}} ^{H}\text{Conv}_{\overline{\mathbf{s}}}(\hat{\mathbf{x}})  \| \\
    &\le 1-\eta+\eta \epsilon\\
    \end{aligned}
\end{equation}
where first equality is due to the fact that $\mathcal{P}(\cdot)=(I-M_{\Omega})\cdot + \mathbf{y}_{\Omega} $ with $\|I-M_{\Omega}\|\le 1$ and the second equality is due to $I=(1-\eta)I+\eta I$. 
This means that $\mathcal{P}(\mathcal{G}_{\mathbf{s}}(\cdot)$ is nonexpansive, i.e.  
\begin{equation}
    \|\mathcal{P}(\mathcal{G}_{\mathbf{s}}(\hat{\mathbf{x}}) ) - \mathcal{P}(\mathcal{G}_{\mathbf{s}}(\hat{\mathbf{x}}') )\| 
     < (\le 1-\eta+\eta \epsilon)\| \hat{\mathbf{x}} - \hat{\mathbf{x}}'\|.
\end{equation}
Then, by the fixed-point theorem, we know that algorithm (\ref{eq:7}) converges to a fixed point $\hat{\mathbf{x}}^{\infty}$.

\end{proof}

\bibliographystyle{ieeetr}
\bibliography{refs}

\end{document}